\documentclass[usenatbib,fleqn]{mn2e}
\usepackage{amsmath}
\usepackage{epsfig}
\usepackage{losymbol}

\newcommand{\apj}{ApJ}
\newcommand{\mnras}{MNRAS}
\newcommand{\aap}{A\&A}
\newcommand{\apjs}{ApJS}

\newcommand{\nature}{Nat}
\newcommand{\apjl}{ApJL}
\newcommand{\aaps}{A\&AS}
\newcommand{\aj}{AJ}

\title[The very bright SCUBA galaxy count]{The very bright SCUBA galaxy count:  looking for SCUBA galaxies with the Mexican Hat Wavelet}

\author[V.E.~Barnard et al.]
{V.E.~Barnard$^{1}$, 
P.~Vielva$^{2}$,
D.P.I.~Pierce-Price$^1$,
A.W.~Blain$^3$,
R.B.~Barreiro$^2$,\newauthor
J.S.~Richer$^4$,
C.~Qualtrough$^4$
\\
\\
$^1$Joint Astronomy Centre, 660 N. A`oh\={o}k\={u} Place, Hilo, HI 96720, USA\\
$^2$Instituto de Fisica de Cantabria (CSIC-UC) Avda. Los Castros s/n, 39005 Santander, Spain\\
$^3$Caltech Astronomy, Caltech 105-24, Pasadena, CA 91125, USA\\
$^4$Cavendish Astrophysics, University of Cambridge, Cambridge, CB3 0HE, UK \\
}

\begin{document}
\maketitle

\begin{abstract}
We present the results of a search for bright high-redshift galaxies in two large SCUBA scan-maps of Galactic regions.  A Mexican Hat Wavelet technique was used to locate point sources in these maps, which suffer high foreground contamination as well as typical scan-map noise signatures.  A catalogue of point source objects was selected and observed again in the submillimetre continuum, and in $\rm{HCO}^{+}$(3$\rightarrow$2) at zero redshift to rule out Galactic sources.  No extragalactic sources were found.  Simulations show that the survey was sensitive to sources with fluxes $\ga 50$ mJy, depending on the local background.  These simulations result in upper limits on the 850-$\mu\rm{m}$ counts of SCUBA galaxies of 53 per square degree at 50 mJy and 2.9 per square degree at 100 mJy.

\end{abstract}

\begin{keywords}
techniques: image processing --- galaxies: evolution --- sub-mm: galaxies

\end{keywords}

\section{Introduction}
\label{sec:intro}

\begin{figure}
\includegraphics[width=150pt,angle=270]{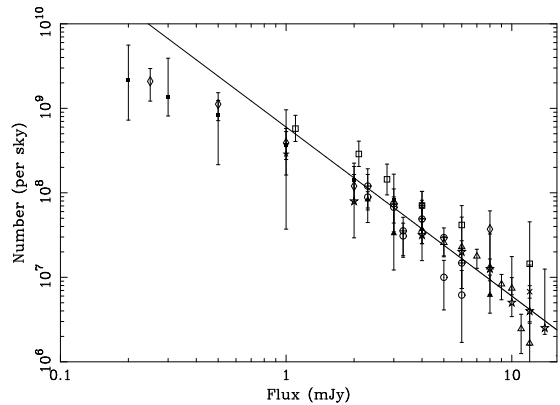}
\caption[Blank field integral counts of SCUBA galaxies]{The integral counts of SCUBA galaxies as measured by the surveys listed in Table \ref{table:blank}.  The solid line shows a power law fitted to the bright count end so that $N \propto S^{-2}$ where $N$ is the number of galaxies per sky and $S$ is the $850 \mu \rm{m}$ flux.}
\label{fig:blankcounts}
\end{figure}

The Submillimetre Common-User Bolometer Array (SCUBA, \citealt{wayne}) has been operating at the James Clerk Maxwell Telescope (JCMT) since 1997, and in that time many blank-field extragalactic surveys have been completed, as summarised in Table \ref{table:blank} \citep{sib,s38,h9,b18,b19,bsikfrayer,e5,b35,cdalens,cbk,s38,s17,serjeant,w11,borys_super}.  From these surveys, the counts of SCUBA galaxies over the flux range 2--15 mJy are now well measured.  This is illustrated in Fig. \ref{fig:blankcounts}, where the counts from several surveys have been compiled.  

This range of fluxes is imposed both by the abilities of SCUBA and the JCMT, and by the nature of the counts.  SCUBA has a beam at 850 $\mu \rm{m}$ of about 14\arcsec, which, in combination with the expected slope of the counts and the clustering strength of these objects, results in a confusion limit for reliable detections of about 2 mJy \citep{confusionII,hogg,s38}.  

\begin{table*}
\caption[Blank field surveys using SCUBA]{A summary of the blank field surveys carried out using SCUBA.  Lensing survey sensitivities have been corrected for lensing.  Symbol in the fifth column refers to that used in Fig. \ref{fig:blankcounts}.}
\begin{tabular}{lcclc}\hline
Survey & Depth (3$\sigma$, mJy) & Area (square arcmin) & Reference & Symbol\\\hline
Hawaiian deep fields & 3 & 9 & \citealt{b18} & Filled triangle \\
& 2 & 7.7 & `` & `` \\
& 8 & 104 & `` & `` \\
HDF jiggle-map & 1.5 & 6 & \citealt{h9} & Filled star \\
CUDSS & 3.5 & 50 & \citealt{e5} & Crossed circle \\
& 3 & 60 & \citealt{w11} & Empty circle \\
HDF widefield scan-map & 9 & 125 & \citealt{b35} & Cross \\
Canada Lens Survey & 1.5 & 19 & \citealt{cdalens} & Empty square \\
Hawaiian Lens Survey & 2.4 & 19 & \citealt{cbk} & Filled square \\
8 mJy survey & 7.5 & 260 & \citealt{s17} & Empty triangle \\
UK SCUBA Lens Survey & 2 & 15 & \citealt{s38} & Empty diamond \\
HDF Super-map & (Varied) & 165 & \citealt{borys_super} & Empty star \\
\hline
\end{tabular}
\label{table:blank}
\end{table*}

At the higher flux end, the limit of $\sim$ 15 mJy is set by the steep slope of the counts, and by the sensitivity of SCUBA.  The variation of the integral galaxy count per unit area, $N$, with limiting flux density, $S$, can be described as a power law, $N \propto S^{-\alpha}$.  Between 5 and 15 mJy, $\alpha$ has been constrained at $\simeq$ 2 (Fig. \ref{fig:blankcounts}).  Beyond about 15 mJy the counts are not well known, although at some brighter flux they are expected to tend to the low-redshift Euclidean slope with $\alpha = 1.5$.  The recent detection of a bright submillimetre quasar, with $S_{850 \mu \rm{m}} \sim 25 $ mJy \citep{kirsten_recent}, strengthens the idea that there is some bright, as yet undiscovered, population of SCUBA galaxies. 

In this paper, we address the question of the very bright count by searching for serendipitous observations of external galaxies in observations of Galactic fields.  Two of the largest SCUBA maps made so far are of the Galactic centre and the Perseus star-forming regions (\citealt{p5}, Qualtrough et al., in prep.), which total approximately 5 square degrees.  SCUBA scan-maps pointed towards known Galactic features are the only SCUBA maps on such large scales.  Whilst each map contains bright Galactic emission, this is of course optically thin in the submillimetre and hence any background extragalactic objects will shine through.  Hence the challenge is to separate the two astronomical components.  Furthermore, SCUBA scan-maps are prone to complicated noise structures, which arise both from the usual atmosphere, telescope and instrument sources and from the image reconstruction technique.  Thus a source extraction routine which does not rely on assumptions about the characteristics of the noise (astronomical and otherwise) in a map is preferable.  For this purpose, a Mexican Hat Wavelet (MHW) algorithm was used.

\section{Scan-mapping with SCUBA: the data}

\label{scan-maps}

The scan-mapping mode available for SCUBA observations is the principal way to make large, shallow maps with SCUBA.  Relatively rarely used in comparison to the jiggle-map technique, scan-maps thus far have been predominantly of Galactic regions.  The only completely blank-field scan-map is the survey of the Hubble Deep Field by \citet{b35}.

In scan-map mode \citep{timscan}, the SCUBA bolometer array is scanned across the sky to cover an area of approximately 10\arcmin $\times$ 10\arcmin, much larger than the single-pointing 2.3\arcmin diameter field of view.  Whilst scanning, a `chop' movement at a rate of $\sim 8$ Hz is used to provide a reference measurement of atmospheric emission.  The secondary mirror of the JCMT switches the array between two patches of sky separated by the chop throw, which is sufficiently small to enable accurate removal of near-field sky emission for objects smaller than the chop throw scale, such as the point sources we are interested in. 

SCUBA scans each field six times, using three chop throws each in two chop
directions, chosen so that identical chop throws are
at $90^{\circ}$ to each other.  The six scans are then added
together by sensitivity-weighted averaging within Fourier space to produce the final
image.  However, the coverage of the Fourier frequency plane is imperfect, resulting
most noticeably in low sensitivity of the array to the larger spatial
scales within the overall scan field.  The 14\arcsec beam size of the telescope also
acts to reduce sensitivity at very small spatial scales.

Prior to the full combination each scan goes through several reduction steps, common to most SCUBA data reduction.  The maps are flatfielded to correct for different sensitivities between bolometers, the extinction of the signal due to the sky is corrected for, spikes caused by cosmic rays are removed, and remaining instrumental baselines and particularly noisy bolometers are removed.  Finally, the time series data is rebinned onto an image grid, here using a pixel scale of 3\arcsec at 850 $\mu \rm{m}$, and the rebinned maps are calibrated using maps of planets or other well-known objects taken in the same observing mode.  These steps were all performed using the SCUBA User Reduction Facility software (\textsc{SURF}, \citealt{surf}).

In the rest of this paper, the term `background' will be used to describe every contribution to the observed flux in images except the point sources i.e. the `background' includes the diffuse Galactic emission, the atmosphere, instrument and telescope noise, and any artifacts such as noise spikes or features related to the reconstruction of the maps in Fourier space.

\section{Point source extraction}
\subsection{\textsc{SExtractor}}
\label{sec:SExtractor}
Our initial attempts to locate point sources in these maps used the Starlink source extraction package known as \textsc{SExtractor} \citep{bertin}.  It is intended primarily for optical images, with perhaps thousands of objects which may be slightly or completely resolved. \textsc{SExtractor} thus identifies sources as clusters of pixels all with flux above some threshold, which is chosen with respect to an estimated background.  Filters can be used to guide the required shape of the clustered pixels.  

Initial runs with \textsc{SExtractor} found that in the crowded Galactic centre fields it was an unreliable detector of even the most obvious point sources:  any source detection software which relies on being able to accurately characterise the background will not perform well with the complicated backgrounds found in SCUBA scan-maps.  As explained below, wavelet routines do not need prior assumptions about the background emission in a map in order to find point sources, and so this detection technique was chosen.  A further advantage of the MHW technique is that it can be finely tuned to finding sources of the appropriate size, leading to fewer false positives.

\subsection{The Mexican Hat Wavelet}
\label{sec:beamshape}

The MHW technique has become a commonly-used technique to separate the point sources from future \emph{Planck} data \citep{c3,s44,v6,v03}.  The combination of the MHW technique with the Maximum Entropy method \citep{hobson} has proved to be very successful in separating all the components of the simulated microwave sky \citep{v5}.

Other wavelet techniques have been suggested for extraction of point sources in CMB maps \citep{tenorio}, but \citet{s44} demonstrated that for Gaussian sources the best wavelet basis to use for a wide range of backgrounds is the Mexican Hat.  This was also found to be preferable to a straightforward Gaussian filter for all backgrounds except the (unrealistic) case of white noise.  A Gaussian linear filter generally produces a smoothing effect, lowering the amplitudes of detected sources with respect to the noise levels within a map.

A wavelet filter acts as a band-pass filter on the Fourier transform (FT) of an image -- depending on the definition of the `mother wavelet' (see section \ref{sec:MHWmaths}), the power from certain spatial frequencies is amplified.  In CMB maps, the removal of several types of foreground signal, including extragalactic point sources, is essential to separate out the cosmological signal.  In the application of the wavelet algorithms in this paper, the extragalactic objects become the final target.  The MHW technique was also used to detect point sources in X-ray data by \citet{damiani}.  We have also applied the MHW technique to a jiggle-map survey of cluster lenses.  The details of the method, which differs slightly from that used here for scan-maps, will be described in K. K. Knudsen et al., in prep. 

\begin{figure}

\includegraphics[width=250pt]{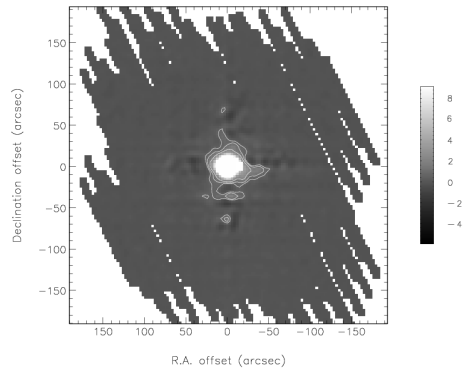}

\caption[Scan-map beam images at 850 $\mu \rm{m}$.]{Scan-map images of the point source calibrator Uranus at 850 $\mu \rm{m}$ as observed in 1999, coeval with the scan-map images of the Galactic centre used in this paper.  The contours begin at 4 Jy $\rm{beam}^{-1}$.  }
\label{fig:scan_calib_850}
\end{figure}

In Fig. \ref{fig:scan_calib_850}, a scan-map image of the point source calibrator Uranus at 850 $\mu \rm{m}$ is shown.  The contours demonstrate that the Gaussian peak of the beamshape is affected by artifacts caused by the incomplete and variable coverage of scanning of the Fourier plane as well as imperfections in the surface of the JCMT's dish.  These artifacts are known to be more prevalent in 450-$\mu \rm{m}$ SCUBA maps, and so the search for point source candidates was carried out using the 850-$\mu \rm{m}$ maps only.  Furthermore, the ratio of Galactic flux to typical (anticipated) extragalactic point source flux is lower at 850 $\mu \rm{m}$, making this the more sensitive wavelength.

\section{Mathematics  and Application of the Mexican Hat Wavelet}
\subsection{Mathematics}
\label{sec:MHWmaths}
In this section the standard mathematics involved in the Mexican Hat transform will be described.  Further detail can be found in the original papers describing this work \citep{c3,v6}.

The continuous isotropic wavelet transform of a 2-dimensional signal $f({\mathbfit x})$ is

\begin{equation}\label{eq:wt}
w_f(R,\mathbfit{b}) = \int d^2\mathbfit{x}\,
\frac{1}{R^2}\psi\left(\frac{|\mathbfit{x} - \mathbfit{b}|}{R}\right)
\,f(\mathbfit{x}),
\end{equation}
where $w_f(R,{\mathbfit b})$ is the wavelet coefficient associated with the scale $R$ at the point ${\mathbfit b}$.  $\psi\left(\frac{|\mathbfit {x}- \mathbfit{b}|}{R}\right)$ is known as the `mother wavelet' which defines the spatial frequencies of interest.  The two-dimensional Mexican Hat mother wavelet is given by 

\begin{equation}\label{eq:mw}
\psi(x) = \frac{1}{\sqrt {2\pi}}\left[2-\left(\frac{x}{R}\right)^2\right]e^{-x^2/2R^2}.
\end{equation}

A Gaussian point source, generated by the convolution of a delta function of amplitude $A$ with a telescope beam of FWHM $\theta$, can be described as 

\begin{equation}\label{eq:ps}
f({\mathbfit x}) = \frac{A}{\Omega} e^{-{\mathbfit x}^2/2\theta^2},
\end{equation}
where $\Omega$ is the area under the beam.  For such an input signal, the wavelet coefficient for a scale $R$ at the position of the source is

\begin{equation}\label{eq:pswc}
\frac{w_f(R,0)}{R} = 2\sqrt{2\pi}\frac{A}{\Omega}\frac{(R/\theta)^2}{[1+ (R/\theta)^2]^2}.
\end{equation}

The effectiveness of the Mexican Hat Wavelet is revealed through the definition of the \emph{detectability} or detection level of the source.  In real space, the criterion for detection of a source is usually the signal-to-noise of the source, given by

\begin{equation}\label{eq:drs}
D_r = \frac{A/\Omega}{\sigma_n},
\end{equation}
where $\sigma_n$ is the dispersion of the real space noise.

In wavelet space, the equivalent parameter for the same point source (as a function of scale $R$) is defined as

\begin{equation}\label{eq:dws}
D_w(R) = \frac{w_f(R,0)}{\sigma_{w_n}(R)},
\end{equation}
where $\sigma_{w_n}(R)$ is the dispersion of the noise field in wavelet space.

Looking at the ratio of these two detectabilities, the amplification, $\mathcal{A}$, due to the transform into wavelet space is

\begin{equation}
\mathcal{A} = \frac{D_w(R)}{D_r} = 2\sqrt{2\pi}R\frac{(R/\theta)^2}{[1+ (R/\theta)^2]^2}\frac{\sigma_n}{\sigma_{w_n}(R)}.
\end{equation}

Thus, for a suitable scale $R_{\rm{opt}} \simeq \theta$, the amplification possible by transforming an image into wavelet space can be very large, because at this scale the dispersion of the general background in wavelet space will be lower than in real space.  Unless the background has Gaussian peaks of the same scale as the point sources (i.e.\ unless the background is completely indistinguishable from the real point sources), it will always be relatively less amplified.

\subsection{Candidate selection process}
\label{sec:detpars}

\begin{figure}
\begin{center}
\includegraphics[width=250pt]{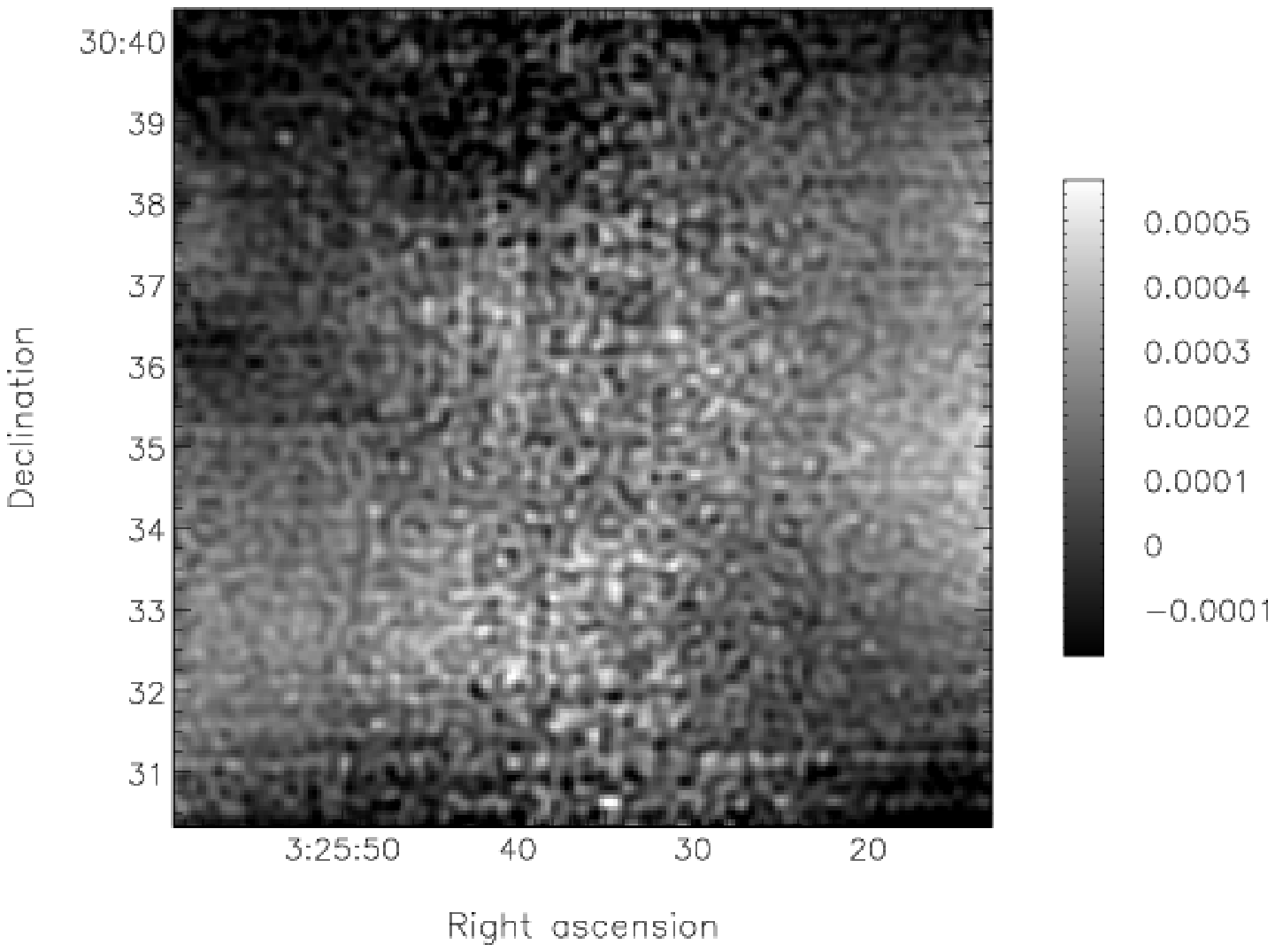}
\includegraphics[width=250pt]{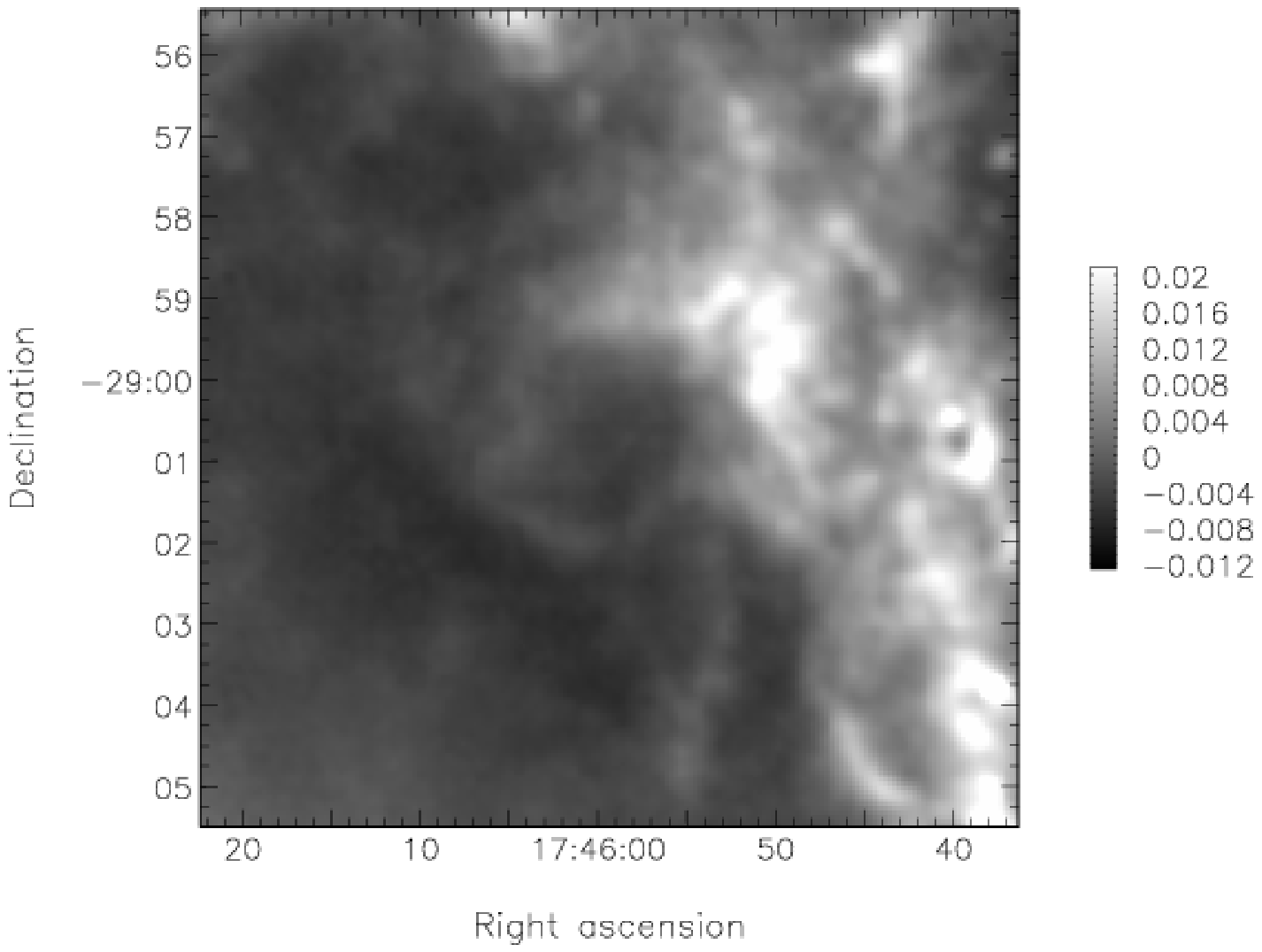}
\end{center}
\caption[Typical squares used for point source detection]{Two squares used as input to MHW point source detection.  The top square is a Perseus field, whilst the bottom square is a Galactic centre field.  These two squares demonstrate the variety of backgrounds that the MHW algorithms had to cope with.  The small-scale (noise) structure visible in the Perseus image is also present in the Galactic centre data, but is masked by the much stronger Galactic emission.}
\label{fig:square}
\end{figure}

As can be seen from Fig. \ref{fig:scan_calib_850}, the scan-map technique produces very irregularly shaped maps.  However, the MHW routines require square input images with no blank pixels.  These were created by trimming each 10\arcmin $\times$ 10\arcmin map individually to keep the largest possible areas available for searching for point sources.  In fact the original maps overlapped in order to mosaic them into the final large maps, and so the overall area lost by trimming the original irregular maps was very small.  Two typical squares from this stage of the process are shown in Fig. \ref{fig:square}.

Simulations performed using the Mexican Hat wavelet algorithm for use with \emph{Planck} \citep{c3,v6,v03} were used to determine the detection parameters for this study, since the effects of various background components at an identical wavelength were considered.  Detection of a source is controlled by two parameters.  The algorithm for source selection was as follows:

\begin{enumerate}
\item Firstly the optimal scale, $R_{\rm{opt}}$, was calculated by the MHW software for each input map.  This involved iterating through small changes in the value of $R$ around the point source scale $\theta$ (the beamsize) until the maximum amplification $\mathcal{A}$ was found for the map.  The final value of $R_{\rm{opt}}$ depends upon the impact of the background on varying scales \citep{v6} -- a background with a characteristic scale a little larger than $\theta$, for instance, can be most strongly counteracted by a value of $R_{\rm{opt}} < \theta$.  As expected, in the Perseus maps, where the most prominent background component is on small scales (Fig. \ref{fig:square}), the value of $R_{\rm{opt}}$ was typically larger than beam size, with a modal value of $1.2\, \times$ beam size.  Conversely, in the Galactic centre maps, the strong large-scale Galactic emission led to a modal value of $R_{\rm{opt}} = 0.7\, \times$ beam size.
\item Point source candidates were then selected at positions with wavelet coefficient values $w_f(R_{\rm{opt}}) \geq 5\sigma_{w_n}(R_{\rm{opt}})$ i.e.\ $D_w(R_{\rm{opt}}) \geq 5$.
\item For each candidate, the `experimental' $w_f(R)$ was compared to the theoretical variation expected with $R$, as a further check on the source's shape.  A value of $\chi^2$ was calculated between the expected and experimental results and a source was retained as a candidate if $\chi^2 \leq 4$.
\end{enumerate}

The 339 candidates (mostly in the Galactic centre fields) flagged by the MHW package were each examined by eye, to rule out those sources which were obviously associated with strong filamentary Galactic features.  This ruled out the majority of the detected point sources.  Next, the detected position of each source was tested against the \textsc{simbad} online database\footnote{http://simbad.u-strasbg.fr/Simbad} to rule out known Galactic point sources, as well as to check whether any known extragalactic objects could be identified with the submillimetre sources.  None of the latter were located.  Candidates were ruled out from further investigation whenever they were close, with respect to the angular resolution both of SCUBA and other instruments, to known Galactic objects. Typical Galactic objects which could appear point-like to the SCUBA 14\arcsec\, 850-$\mu \rm{m}$ beam include Herbig-Haro objects or protostars, as well as knots of dust within larger structures, especially along filaments \citep{p5,visser}.

From this process a final list of 21 candidates was obtained, 15 in the Perseus region and 6 in the Galactic centre region\footnote{Included in the Galactic centre list was an object (gc6) which, whilst only weakly detected by the MHW routine, had extremely strong fluxes and was apparently point-like at both 850 and 450 $\mu \rm{m}$.  This object was retained because despite its bright flux, no other survey in \textsc{simbad} showed a detection nearby.  Its bright flux indicated that only a very short integration time would be necessary to confirm it as a real source.  The reason for its very weak detection by the wavelets routines despite its shape and strength was not clear at this point (but became clear following later simulations -- see section \ref{sec:losses}).}.  Of the original list, 94\, per cent were rejected either by association with clear Galactic features or by the \textsc{simbad} comparison.  The 21 candidates, with their MHW-estimated fluxes at 850 $\mu \rm{m}$, are listed in table \ref{table:candidates}.

\begin{table*}
\begin{center}
\caption[Point source candidates]{Observations of point source candidates from the Galactic centre and Perseus region maps.  Positions (J2000) and fluxes are those \emph{predicted} by the MHW routines.  The final three columns give observing details: observation date, time spent integrating on each source and average sky opacity during observation. }
\label{table:candidates}
\begin{tabular}{lcccccc}\hline
Source & R.\ A. & Dec. & $S_{850\mu \rm{m}}$ (mJy) & Obs. date & $\rm{T}_{\rm{int}}$ (s) & $\tau_{850 \mu \rm {m}}$ \\\hline
gc1 & 17 46 42.7 & $-28\ 08\ 20$ & 319  & 26/08/01 & 640 & 0.340\\
gc2 & 17 44 19.2 & $-29\ 33\ 35$ & 344 & 26/08/01 & 640 & 0.340\\
gc3 & 17 44 05.1 & $-29\ 41\ 53$ & 224 & 26/08/01 & 1024 & 0.340\\
gc4 & 17 44 13.9 & $-29\ 40\ 47$ & 245 & 26/08/01 & 768 & 0.340\\
gc5 & 17 44 21.4 & $-29\ 49\ 04$ & 232 & 26/08/01 & 1024 & 0.340\\
gc6 & 17 43 10.3 & $-29\ 51\ 44$ & 1556 & 26/08/01 & 384 & 0.340\\\hline
per1 & 03 25 41.9 & +30 40 57 & 83 & 31/08/01 & 1280 & 0.265\\
per2 & 03 25 35.6 & +30 31 27 & 100 & 30/08/01 & 1920 & 0.350\\
per3 & 03 25 34.6 & +30 30 36 & 63 & 31/08/01 & 2560 & 0.275\\
per4 & 03 25 56.5 & +30 39 52 & 143 & 30/08/01 & 1280 & 0.336\\
per5 & 03 26 24.0 & +30 20 28 & 109 & 31/08/01 & 1280 & 0.275\\
per6 & 03 26 51.8 & +30 28 09 & 98 & 31/08/01 & 1280 & 0.330\\
per7 & 03 45 16.9 & +32 04 47 & 137 & 30/08/01 & 1280 & 0.350\\
per8 & 03 28 13.5 & +31 36 24 & 158 & 30/08/01 & 1024 & 0.336\\
per9 & 03 34 54.6 & +31 12 15 & 84 & 31/08/01 & 1280 & 0.330\\
per10 & 03 33 04.2 & +30 51 45 & 101 & 31/08/01 & 1280 & 0.275\\
per11 & 03 32 40.3 & +31 03 03 & 121 & 31/08/01 & 1280 & 0.265\\
per12 & 03 33 01.7 & +30 52 24 & 141 & 30/08/01 & 1280 & 0.358\\
per13 & 03 33 34.1 & +31 20 09 & 136 & 30/08/01 & 1280 & 0.358\\
per14 & 03 36 42.7 & +31 14 40 & 86 & 31/08/01 & 1280 & 0.330\\
per15 & 03 36 40.1 & +31 05 10 & 96 & 31/08/01 & 1280 & 0.275\\\hline
\end{tabular}
\end{center}
\end{table*}

\section{Follow-up observations of the candidate sources}

To verify the point source candidates found by MHW, secondary observations of their locations with SCUBA were carried out in August 2001.  The aim of these observations was to:  i) verify their point source nature, by re-imaging them in jiggle-map mode at both 850 and 450 $\mu \rm{m}$ -- the smaller beam of the 450-$\mu \rm{m}$ observations providing a check that the candidates, if detected at 450 $\mu \rm{m}$, still appeared point-like on smaller scales; ii) determine accurate fluxes and positions for the candidates, in a map with a single pointing at each location.

\subsection{SCUBA jiggle-maps of the point source candidates}
\label{sec:SCUBAjiggle}
During the nights of the 26th, 30th and 31st of August 2001, each of the 21 candidate positions was observed with SCUBA.  Each position was observed for long enough that, based on the estimated fluxes from the MHW output, the candidates should be detected at more than 5-$\sigma$ levels in both wavebands.  The 450-$\mu \rm{m}$ fluxes were estimated assuming a spectral index in the submillimetre of $\beta \sim 1.5$, where flux $S_{\nu} \propto \nu^{\beta}$.  Details of these observations are listed in table \ref{table:candidates}.

Standard SCUBA jiggle-maps produce a single 2.3\arcmin\, field, centred on the candidate position.  This mode is so-called because during observations the array is `jiggled' to ensure the entire image is fully sampled.  This is unnecessary in scan-map mode as the array moves across the sky and so fills in the gaps between bolometers.  In jiggle-map mode, as well as chopping, an extra `nod' movement is included.  After chopping between the on-source position and an off-source position to one side of the source, the telescope nods so that the secondary mirror now chops  between the on-source position and a new off-source position, at the same distance but on the opposite side of the on-source position.  The main advantage of using the nod and chop combination is that an estimate of any overall gradient in the sky emission can be made.  The chop throws and directions were chosen individually for each source to avoid chopping onto particularly strong areas of Galactic emission visible in the scan-maps.

All of the maps from the jiggle-map observations were reduced using standard \textsc{Surf} reduction routines (\citealt{surf}), following the same steps as described above for scan-maps: flatfielding, extinction correction, spike removal, noisy bolometer removal, and rebinning.  Finally all were calibrated using planets or other well-known point sources.  

\begin{figure*}
\includegraphics[width=250pt]{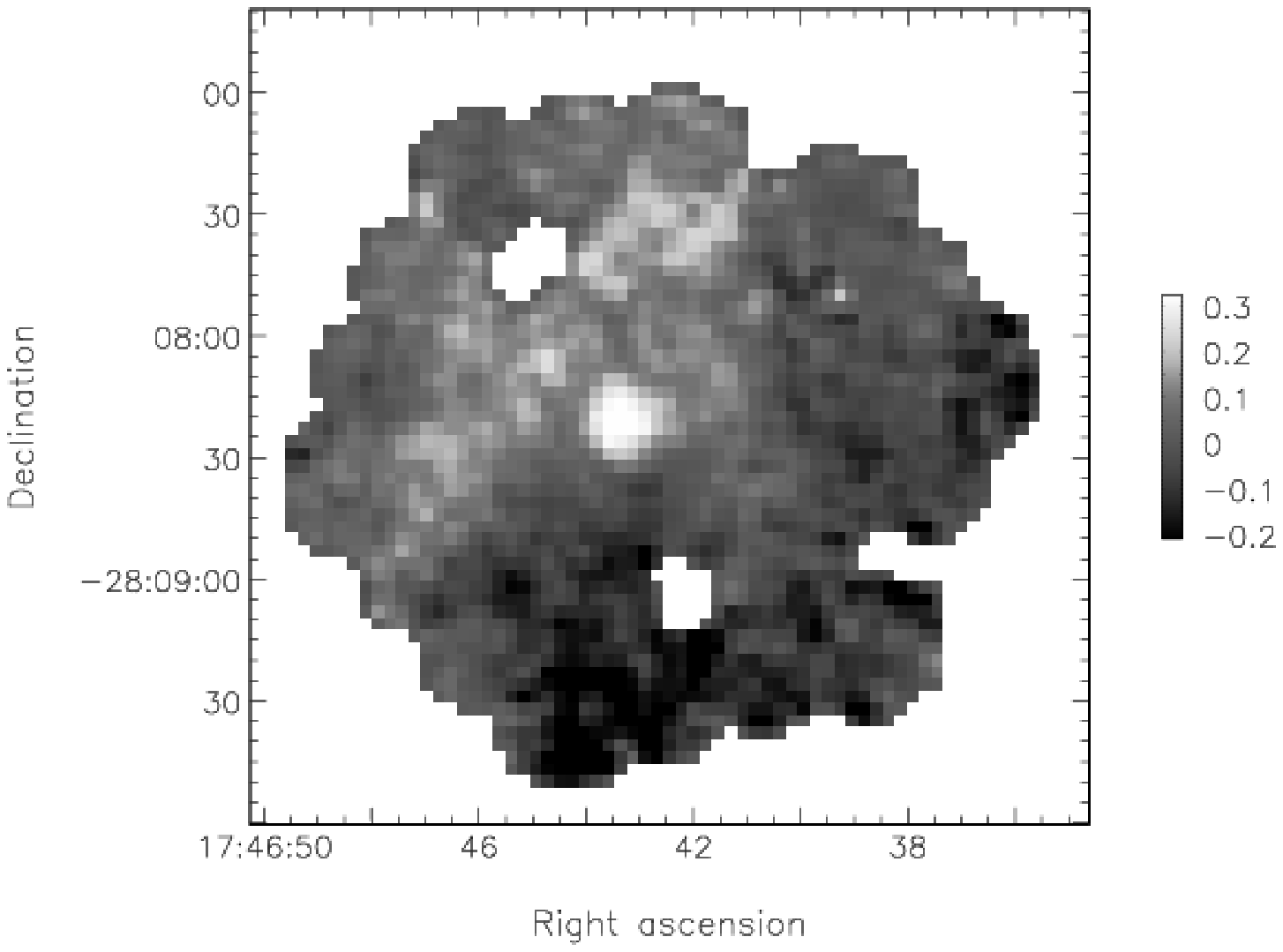}
\includegraphics[width=250pt]{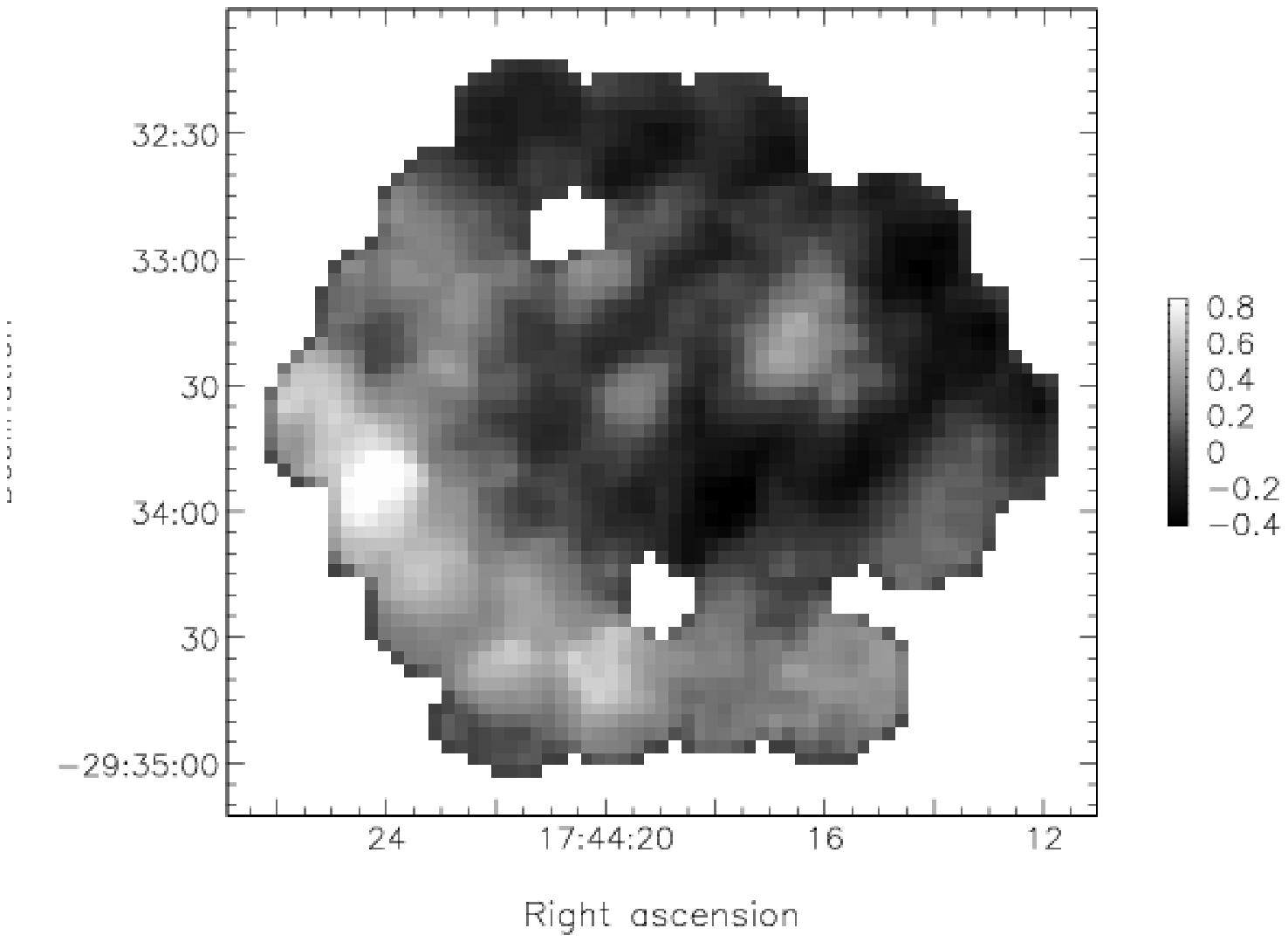}

\includegraphics[width=250pt]{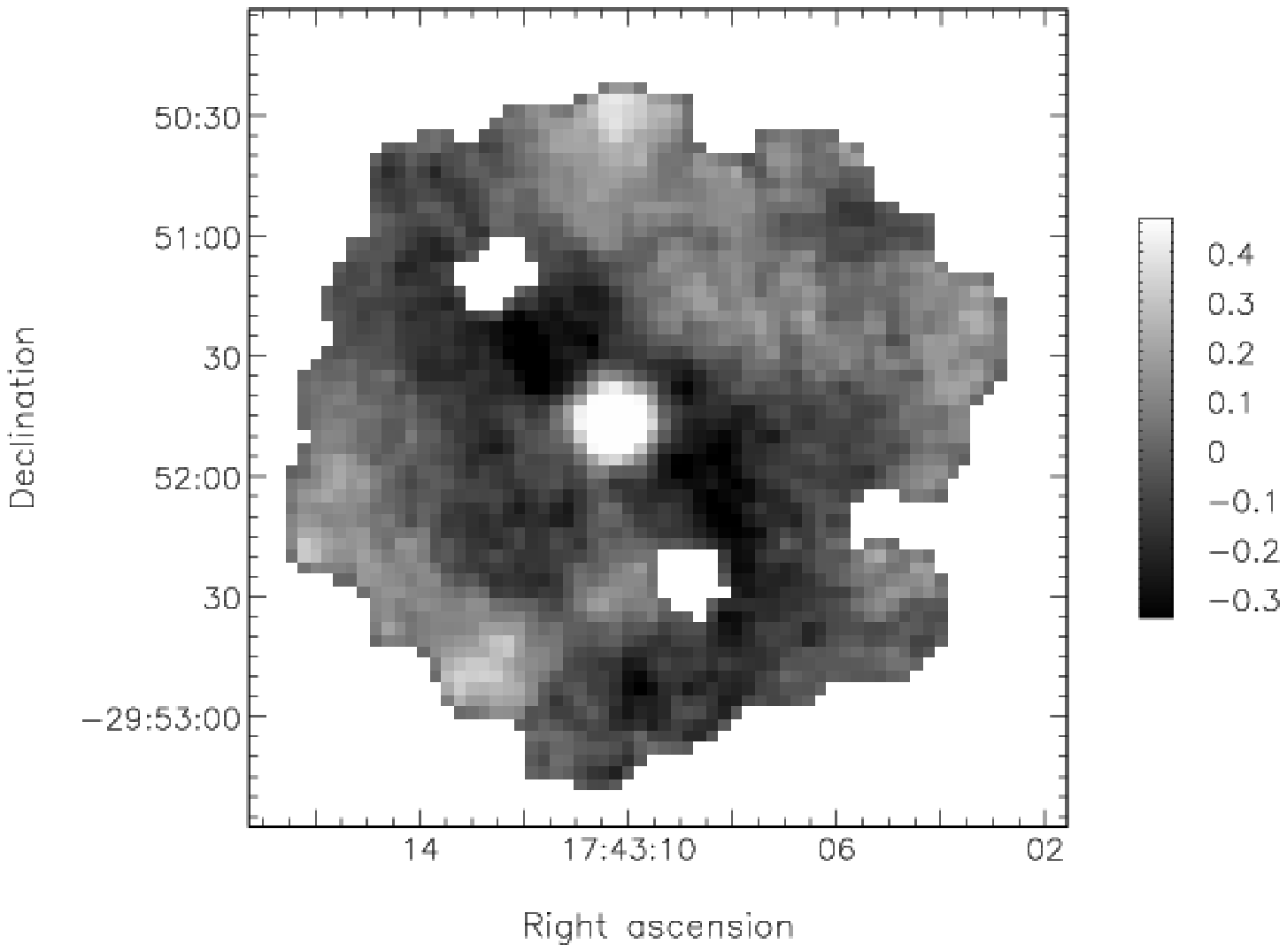}
\includegraphics[width=250pt]{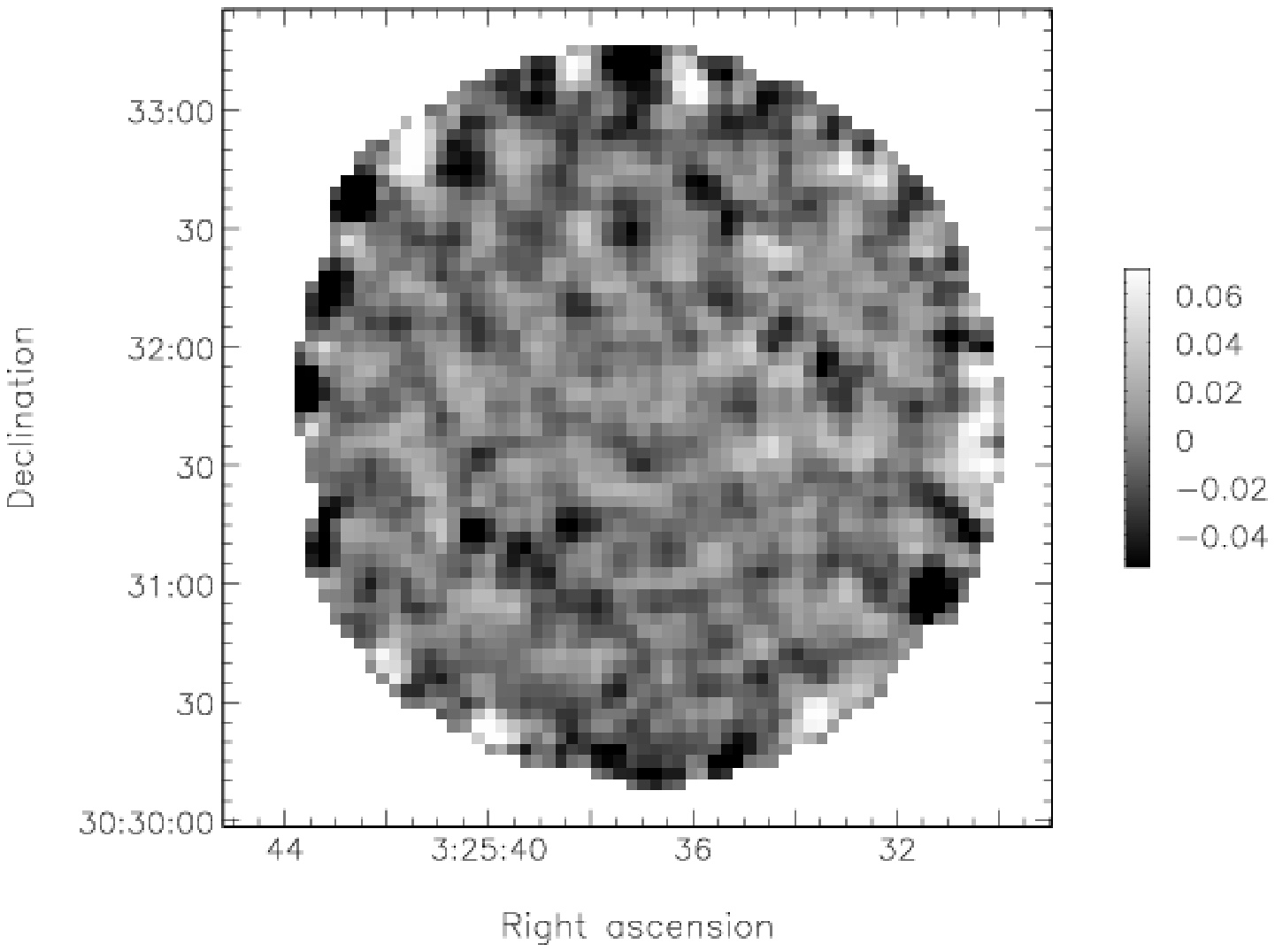}
\caption[850-$\mu \rm{m}$ jiggle-maps of point-source candidates ]{850-$\mu \rm{m}$ jiggle-maps of point-source candidates.  Three maps show the only 850-$\mu \rm{m}$ detections (top left -- object gc1; top right -- gc2; bottom left -- gc6); the other (object per3) is included as an example of the non-detections.  Units are $\rm{Jy\, beam}^{-1}$.  The uneven edges and holes are caused by `turning off' noisy bolometers.}
\label{fig:SCUBA_850_candidates}
\end{figure*}

\begin{figure}
\includegraphics[width=250pt]{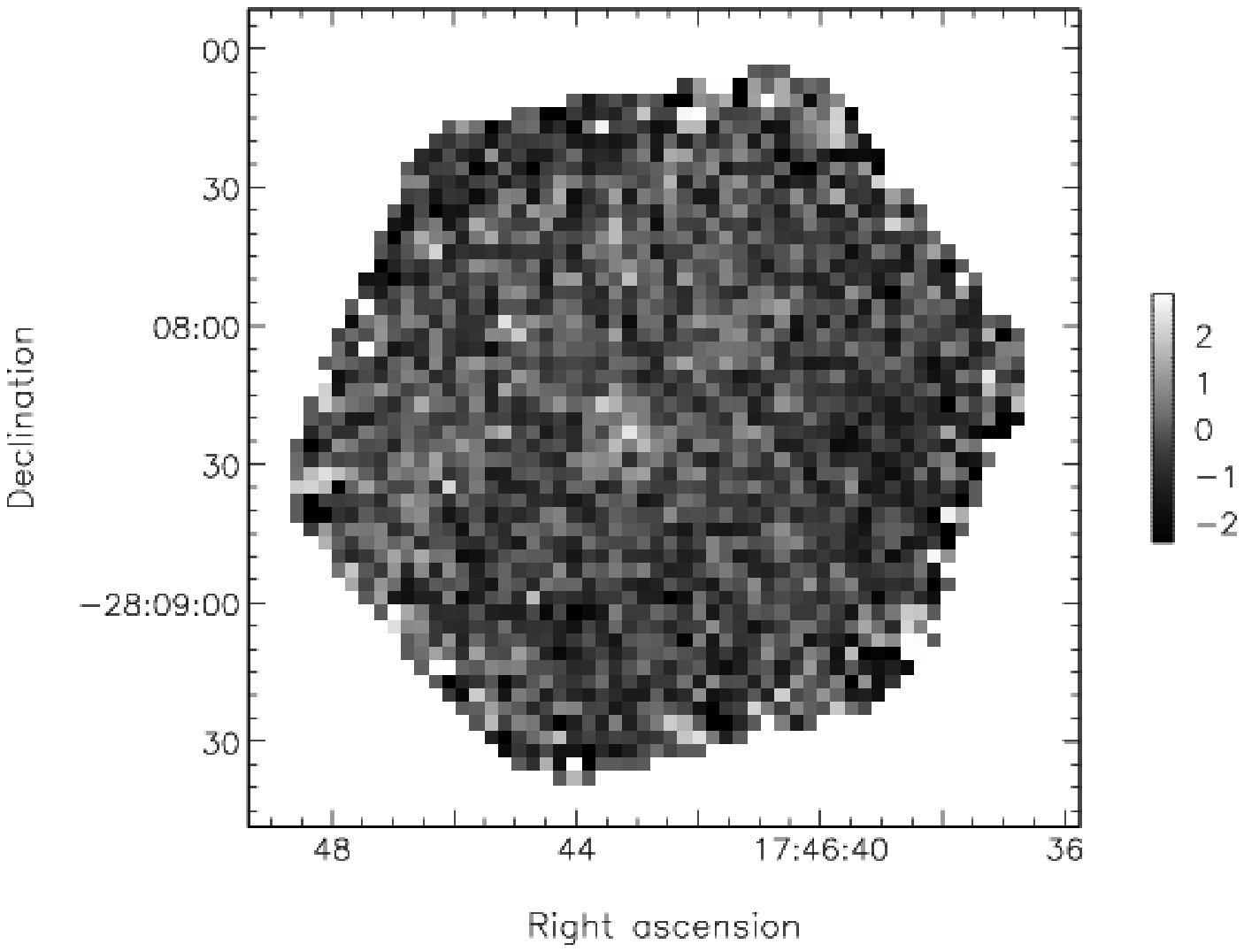}
\includegraphics[width=250pt]{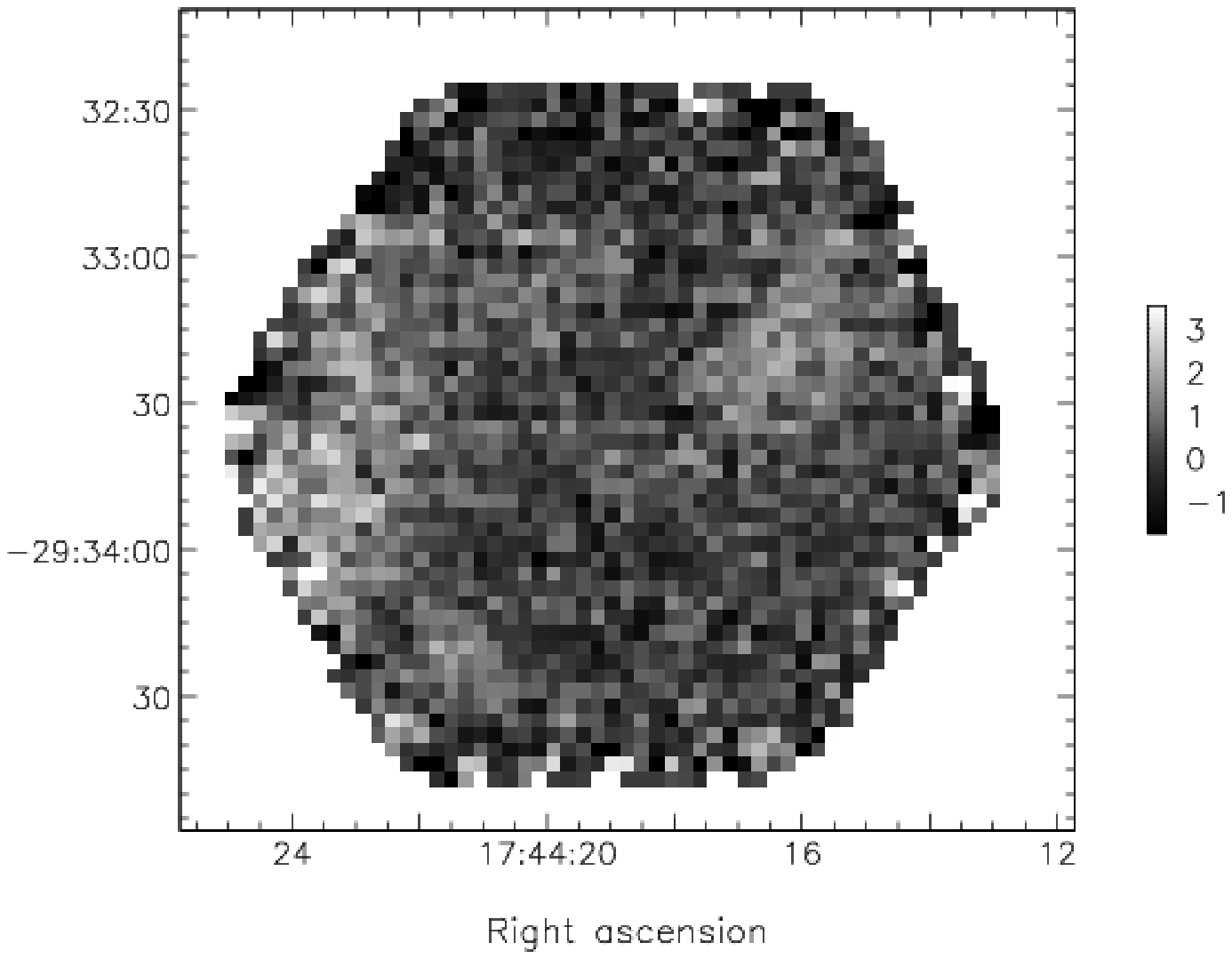}
\includegraphics[width=250pt]{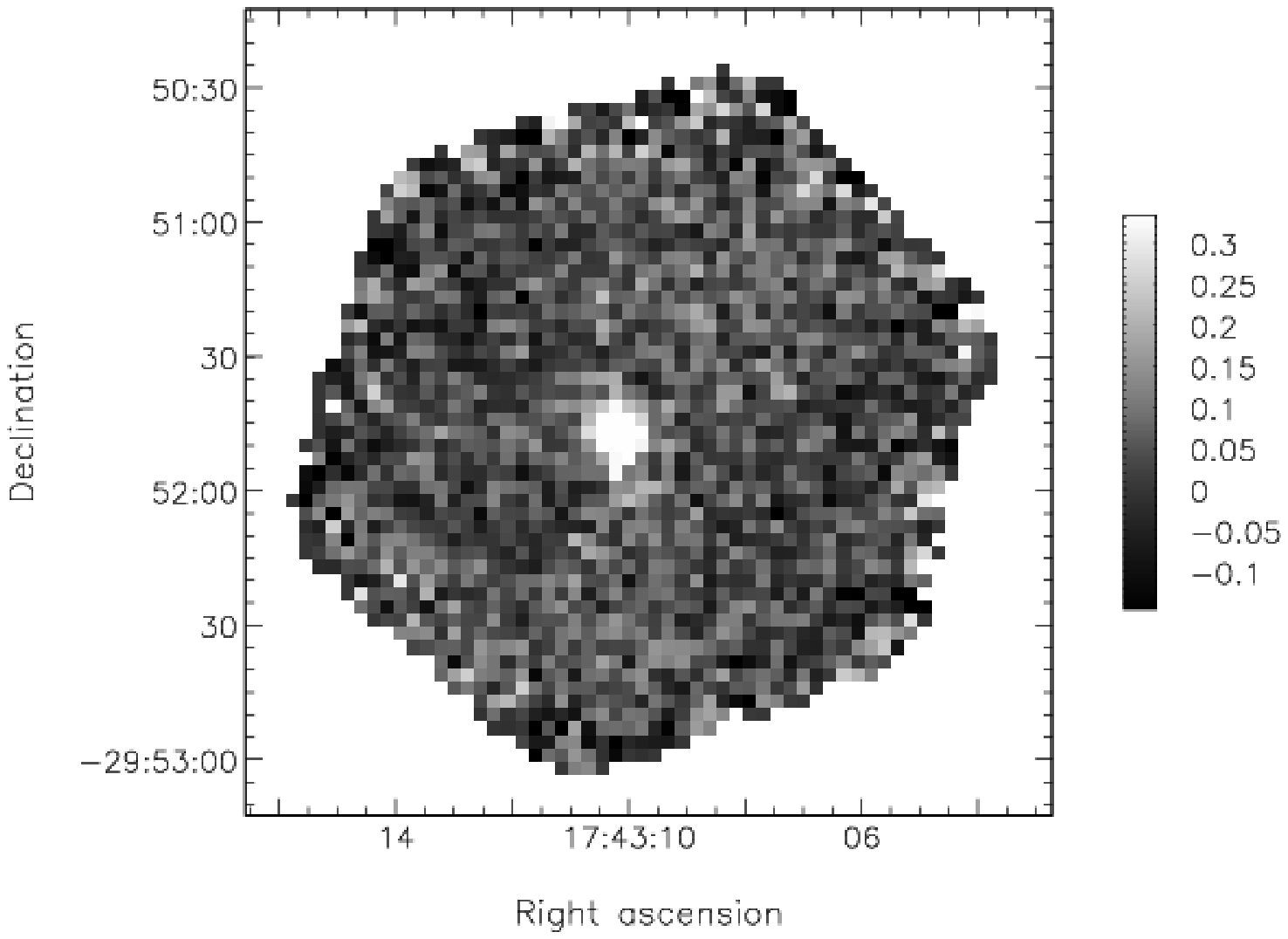}
\caption[450-$\mu \rm{m}$ jiggle-maps of the good point source candidates]{450-$\mu \rm{m}$ jiggle-maps of gc1, gc2 and gc6.  Units are Jy beam$^{-1}$. }
\label{fig:SCUBA_450_candidates}
\end{figure}

The success rate of locating the point source candidates was low.  Three maps, corresponding to the sources labelled gc1, gc2 and gc6 in Table \ref{table:candidates}, do show point source objects at their centre, whereas the remainder show either much larger-scale structure or nothing at all.  The maps of these three sources are shown in Fig. \ref{fig:SCUBA_850_candidates}, along with an example of an empty field where the candidate clearly is absent.  The 450-$\mu \rm{m}$ data for the three detected objects are shown in Fig. \ref{fig:SCUBA_450_candidates}.  The brightest source at 850 $\mu \rm{m}$, gc6, can be seen in the higher frequency map, but the other two cannot.  None of the other candidates had 450-$\mu \rm{m}$ detections of a point-source nature.  The details of the detections of these three objects in the SCUBA maps are given in table \ref{table:SCUBAdets}.

\begin{table*}
\begin{center}
\caption[Secondary observation detections]{Secondary observation detections.  Positions are the peak of the emission in the jiggle-maps; note the good agreement with the positions predicted by MHW in table \ref{table:candidates}.  Fluxes here are the peak values of the point sources taken directly from the jiggle-maps, with 1-$\sigma$ noise levels estimated according to the SCUBA integration time calculator - the backgrounds are too high in these maps to determine noise directly.  Fluxes for gc1 and gc2 have 3-$\sigma$ upper limits only at 450 $\mu \rm{m}$.  For gc1 and gc2, the spectral index $\gamma$ is evaluated for this 3-$\sigma$ value (where $f \propto \nu^{\gamma}$). For gc6, the spectral index is calculated for the measured values of flux at both wavelengths.}
\label{table:SCUBAdets}
\begin{tabular}{l|cc|c|c|c}\hline
Source & R.A. & Dec. & 850-$\mu \rm{m}$ flux (mJy) & 450-$\mu \rm{m}$ flux (mJy) & $\gamma$ \\\hline
gc1 & 17 46 43.5 & -28 08 21.1 & 335 $\pm$ 20 & $\leq 1500$ & $\leq 2.3 $\\
gc2 & 17 44 19.5 & -29 33 33.1 & 100 $\pm$ 20 & $\leq 1500$ & $\leq 4.2$\\
gc6 & 17 43 10.2 & -29 51 47.0 & 1069 $\pm$ 20 & 11318 $\pm$ 600 & 3.7\\\hline
\end{tabular}
\end{center}
\end{table*}

\subsection{Molecular line observations}

To verify whether the three detected point source objects could be extragalactic, they were observed using the heterodyne Receiver A3 on the JCMT, at the Galactic centre rest frequency of the line HCO$^+$(3$\rightarrow$2).  This line is a useful tracer of very dense molecular gas -- the critical $\rm{H_2}$ density to populate the HCO$^+$(J = 3) state is $\sim 7.5 \times 10^{7}\, \rm{cm^{-3}}$ \citep{wild}, within the range expected for molecular clouds responsible for star formation.  HCO$^+$(3$\rightarrow$2) was used in preference to CO lines as it is less widely distributed in the Galaxy, and so the likelihood of confusion with another Galactic source is very low.  The absence of a Galactic HCO$^+$(3$\rightarrow$2) line in these objects would suggest that an extragalactic origin for the sources could be plausible.  

The observations were carried out in frequency-switched mode, which is analogous to a nod and chop in frequency rather than in space.  The signature of a detection in frequency switched mode is therefore a negative--positive--negative shape in frequency space.  Raster maps were made which pointed at a square of $5 \times 5$ positions and covered an area of 40\arcsec $\times$ 40\arcsec around the MHW-derived positions, to allow for possible positional errors in the predictions.  The resolution of the JCMT at the frequency of HCO$^+$(3$\rightarrow$2) is about 18\arcsec, slightly larger than the SCUBA 850-$\mu \rm{m}$ beam size.

\begin{figure}
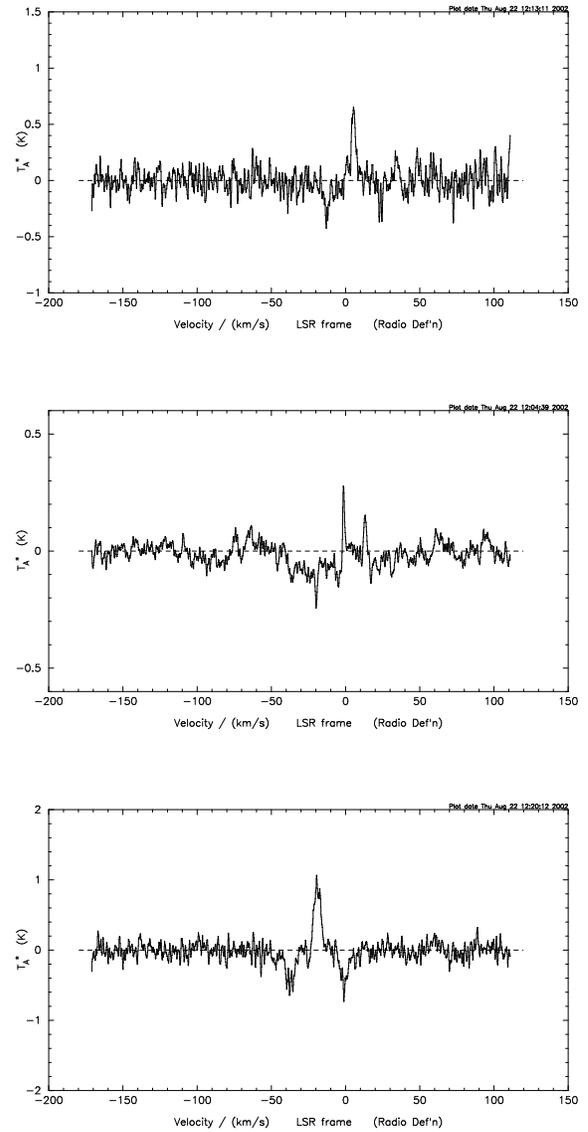

\begin{center}
\includegraphics[clip=,width=150pt,angle=270]{figure6a.ps}
\includegraphics[clip=,width=150pt,angle=270]{figure6b.ps}
\includegraphics[clip=,width=150pt,angle=270]{figure6c.ps}
\end{center}
\caption[Spectra of sources]{Spectra of sources gc1, gc2 and gc6 (top to bottom).  All observations were done on 26/08/01.}
\label{fig:specxobs1}
\end{figure}

The data were reduced using the standard routines from the \textsc{specx} package \citep{specx}.  The results are shown in Fig. \ref{fig:specxobs1}.  Each of the spectra shown have had a polynomial baseline fitted and removed, and have been smoothed into bins using a 5-point running average.

\subsection{Discussion of the point source detections in continuum and lines}

\subsubsection{gc1}

Source gc1 was reasonably strongly detected at 850 $\mu \rm{m}$ in a fairly clear field, but was not detected at 450 $\mu \rm{m}$, although there is a hint of a peak in the right place.  The position and flux of this object in the jiggle-map both agreed well with the MHW predictions.  The central and average molecular line spectra of gc1 both show the expected signature of a dense molecular gas cloud in the Galactic centre (i.e. at $v_{\rm{LSR}}$ $\sim 0\rm{km s^{-1}}$), and so we were able to reject it as an extragalactic candidate.  

\subsubsection{gc2}

Source gc2 was detected in a field with a visibly high background, with an apparent slight extension on the northwest-southeast axis.  Its position again is extremely similar to that predicted, but its peak flux is somewhat below the prediction by MHW.  It was not detected in the 450-$\mu \rm{m}$ map.  The HCO$^+$(3$\rightarrow$2) spectra show a more complicated structure.  The two positive peaks at about 0 and 20 $\rm{km s^{-1}}$ probably represent emission from two clouds at slightly different positions in the Galaxy, sitting on the same line-of-sight.  This may be the cause of the slight extension of the source in the jiggle-map.  The MHW technique is not particularly powerful in separating two closely overlapping point sources, as demonstrated here\footnote{However, if two peaks are known to be overlapping, it is possible to `tune' the detection algorithms to separate the two.  This will be described at more length in K. K. Knudsen et al., in prep.}.

\subsubsection{gc6}

The strongest source in the candidate list, gc6 was detected strongly in the jiggle-maps at both 850 and 450 $\mu \rm{m}$.  Again its position was extremely close to that predicted by MHW, and its 850-$\mu \rm{m}$ peak flux was 30\, per cent lower than predicted.  It is clearly a compact source.  The molecular line observations strongly confine this also to be a Galactic object.

Since both 450- and 850-$\mu \rm{m}$ detections were achieved for this very bright source, it is worth speculating further as to its possible identity.  Fitting a standard blackbody to the 450/850 colours, a plausible value of emissivity $\beta \la 2.0$ is found for temperatures $\ga 80$ K.  However, fitting a single temperature to what is obviously a dense object is probably not realistic.  

A 1.4-GHz VLA survey of the area \citep{liszt,gray} suggests that this position is within an extended structure which has an integrated flux density of several hundred mJy at 1.4 GHz.  However, no compact radio object was identified within 1.3\arcmin of gc6, suggesting an upper limit of 3 mJy at 1.4 GHz.  There is also an \emph{IRAS} object 2\arcmin away, but this also seems unlikely to be the true counterpart.  We are continuing to investigate this unusually bright and compact object.

In summary, the follow-up SCUBA observations demonstrate that of 21 objects detected by the MHW routines, only 3 were genuine point sources.  The heterodyne observations suggest that all 3 of these are Galactic.  The remaining 18 sources appear to be spurious.

\subsection{Understanding the spurious point sources: filtering in scan-maps}

\begin{figure}
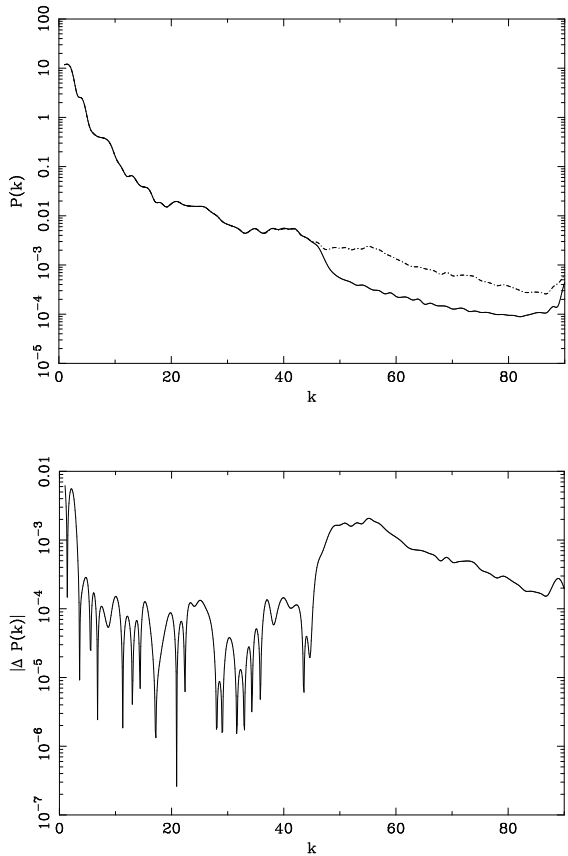

\begin{center}
\includegraphics[width=150pt,angle=270]{figure7a.ps}\vspace{20pt}
\includegraphics[width=150pt,angle=270]{figure7b.ps}
\caption[Power spectra of filtered and unfiltered maps]{Power spectra of filtered and unfiltered maps.  The top plot shows the two power spectra -- the filtered map is represented by the solid line and the unfiltered map by the dot-dash line.  The bottom plot shows the absolute value of difference between the two power spectra.}
\label{fig:filteringexample}
\end{center}
\end{figure}

In order to understand why MHW appeared to find so many spurious sources, we investigated in detail the original scan-map reduction, which revealed an important clue:  in the final data reduction stage, where the six maps are recombined in Fourier space, an option to filter the maps with a low-pass filter had been used in both the Galactic centre and Perseus maps.  The effect of this standard \textsc{Surf} option can be seen in Fig. \ref{fig:filteringexample}, where the power spectra obtained from performing the recombination with and without the filtering option for the same field are shown.

The filtering cuts in most steeply at a scale of $k \ga$ 45, which corresponds to a spatial scale of $\sim \frac{1}{2} \frac{\lambda}{D}\,$, where $\lambda$ is the observing wavelength (850 $\mu \rm{m}$) and $D$ is the diameter of the dish of the JCMT (15 m).  (The 850-$\mu \rm{m}$ beam size of 14\arcsec corresponds to $k \sim 19$.)  However, the difference plot demonstrates that the filtering has an impact at all scales.  The original intention of the filter was to remove noise which appeared on scales smaller than the beam, since this would usually be assumed to be caused by instrumental factors.  However, the information on these small spatial scales is important when attempting to detect point sources.  In particular, noise spikes in a scan-map are smoothed by the filtering to the right spatial scale to create false positives in the MHW algorithms.  Whilst of course the initial data reduction included spike removal, the sensitivity of the MHW detections went to a deeper level than anticipated at that time.  

The maps containing the spurious sources were all re-reduced without the filtering option applied, and in every case the point source was no longer detected, using the same detection parameter requirements.  It appears that at each location, a noise peak was sufficiently altered by the filtering that the MHW routine could not distinguish it from a true point source.  In the unfiltered maps, each spurious source no longer appeared Gaussian to the MHW routine (nor, usually, to the eye), whilst the three real sources were still detected.  

These investigations therefore demonstrate that on unfiltered SCUBA scan-maps, the MHW routines are a reliable detector of point sources.  However the filtering option used as default in many scan-map images of extended sources will mislead detection algorithms into detecting spurious point sources.

\section{Simulations}
\label{sec:sims}
\subsection{Simulating detections of point sources in filtered scan-maps}

The scan-map filtering option made a large impact on the MHW results by creating spurious sources.  However, in the earlier stages of analysis, some candidates detected by MHW in the filtered maps were ruled out because of their alignment with the positions of known Herbig-Haro objects or other sources in the Galaxy.  Also, three sources identified by MHW were confirmed in the later jiggle-maps.  Hence it is clear that \emph{some} real point sources were detectable in the filtered maps.  Thus we have carried out simulations to determine the sensitivity and hence completeness of the analysis already carried out, and what selection effects, if any, the filtering option had on the detectability of the real point source population.     

To evaluate this, simulated point sources were added to each of the Galactic centre and Perseus region squares at known positions and with known fluxes, and the resulting images were fed into the MHW routines following exactly the same procedure as used in the original candidate selection.  The added point sources were spread far enough apart that they did not impact upon each other.  In a data array with the same dimensions as the final map, delta functions of unit flux were placed at chosen positions.  This array was then convolved with a dual Gaussian beam to represent the signature of a point source in the uncombined maps.  This filtering process was repeated six times to produce maps representing each of the combinations of chop throw and direction in the uncombined maps.  At this stage the six simulated source maps were multiplied by a factor to set the flux of the point sources, and then each source map was added to the respective real uncombined map.  These six maps were then recombined, both with and without filtering.  The fluxes of the simulated sources were each gradually increased until the MHW detection of each source occurred.  

\subsubsection{Overall sensitivity}

\begin{figure}
\begin{center}
\includegraphics[width=150pt,angle=270]{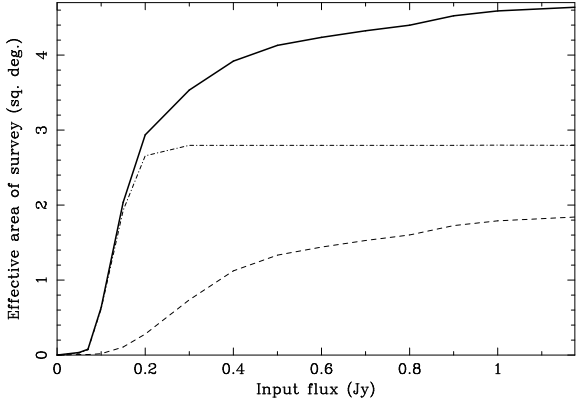}
\end{center}
\caption[Areal sensitivity of the survey]{Areal sensitivity of the survey.  Area within the Perseus region is shown by the dot-dash line, while area within the Galactic centre region is by the dashed line.  The total area is shown by the solid line.  Effective area here is the fraction of sources detected with a given input flux, multiplied by the total survey area of 4.638 square degrees}
\label{fig:area_sens}
\end{figure}

The overall sensitivity of the MHW routines to point sources in the filtered maps of the Galactic centre and the Perseus star-forming region is shown in Fig. \ref{fig:area_sens}.  This figure demonstrates that the sensitivity in the Perseus region was far higher than in the Galactic centre, which is unsurprising considering the higher background flux and more complicated structure in the Galactic centre maps.  In fact the Galactic centre count almost seems to have two slopes, with a break at around 0.4 Jy.  This is probably caused by the variety in the background in the Galactic centre, easily observed by eye (see figures in \citealt{p5}): those fields closer to the Galactic plane suffer from higher backgrounds than those in regions slightly offset from the Galactic plane.  Hence the simulated sources in the squares nearer the Galactic plane can only be detected at higher fluxes. 

\subsubsection{Flux accuracy}
\label{subsubsec:fluxacc}

\begin{figure}
\begin{center}
\includegraphics[width=150pt,angle=270]{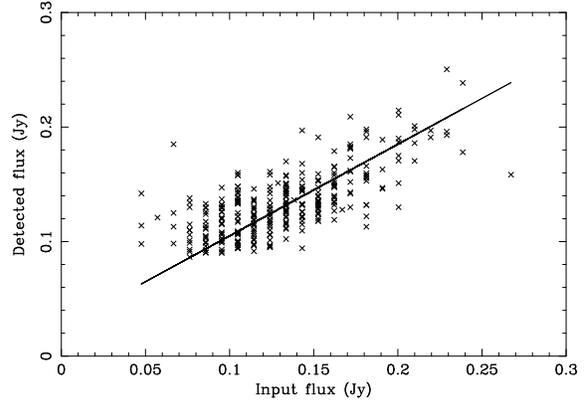}\vspace{20pt}
\includegraphics[width=150pt,angle=270]{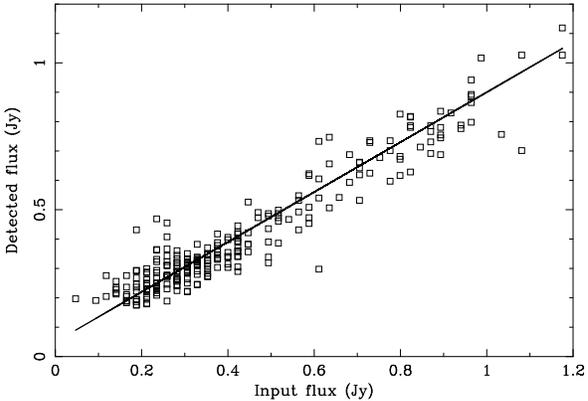}
\end{center}
\caption[Flux accuracy in source detection]{Flux accuracy in source detection.  The quantisation in input flux reflects the resolution of the simulations in terms of input flux values tested.  The upper plot shows the results for the Perseus region fields, while the lower plot shows the Galactic centre results.  }
\label{fig:inp_det}
\end{figure}

In Fig. \ref{fig:inp_det}, the input and output fluxes for simulated point sources in the two survey regions are shown.  Included on each plot is a line fitted by a least-squares method to evaluate the overall gradient and offset.

For a line $ y = mx + c$, the values found for the Perseus region data were $m = 0.80 \pm 0.05, c = 0.025 \pm 0.005$, whilst in the Galactic centre data the appropriate values were $m = 0.85 \pm 0.05, c = 0.050 \pm 0.005$.  Hence it appears that, on average, the true fluxes are generally slightly under-estimated by MHW by 10 or 15\, per cent, although each graph has a positive intercept.  This trend for under-estimation may be due to the imperfect beam shapes of the scan-map sources (see section \ref{sec:beamshape}).  The scan-map reconstruction means that some of the flux of these simulated sources was lost into the artifacts surrounding the central Gaussian.  The MHW technique is not sensitive to these artifacts, and so will underestimate the flux of a source.

\begin{figure}
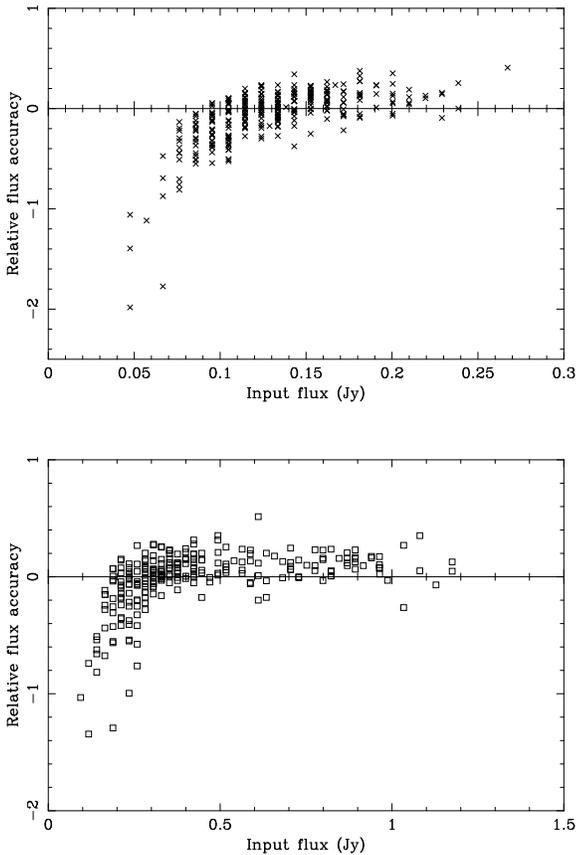

\begin{center}
\includegraphics[width=150pt,angle=270]{figure10a.ps}\vspace{20pt}
\includegraphics[width=150pt,angle=270]{figure10b.ps}
\end{center}
\caption[Relative flux accuracy of source detection]{Relative flux accuracy in source detection. Lines are fitted to distribution -- see section \ref{subsubsec:fluxacc}. Upper and lower plots are as in Fig. \ref{fig:inp_det}.}
\label{fig:fra_inp}
\end{figure}

The absolute scatter in these graphs appears to stay the same at all input fluxes, except perhaps at the very lowest fluxes which appear to be systematically over-estimated. This suggests that the relative accuracy is worse at lower fluxes.  This is confirmed in Fig. \ref{fig:fra_inp}, where the Relative Flux Accuracy (RFA) is shown in comparison with the input fluxes required to detect a source.  Here the RFA is defined as

\begin{equation}
\rm{RFA = \frac{Input\, flux - Detected\, flux}{Input\, flux}}.
\label{RFA}
\end{equation} 

Hence values of RFA $\sim 0$ represent accurate detections, negative values represent over-estimations of flux and positive values of RFA represent under-estimations.

The two graphs show similar trends, indicating that the fluxes of faint sources tend to be over-estimated, whereas brighter fluxes suffer from the effect of the non-Gaussian beam as discussed above.  At low fluxes this can be understood as the impact of the noise in the region of the point source -- a faint source is more likely to be detected if it sits in a positive noise region.  Whilst the MHW routines will not be affected by noise on scales much larger than the beam, peaks or holes on scales similar to the beam can affect the detection accuracy.  

\subsubsection{Positional accuracy}

\begin{figure}
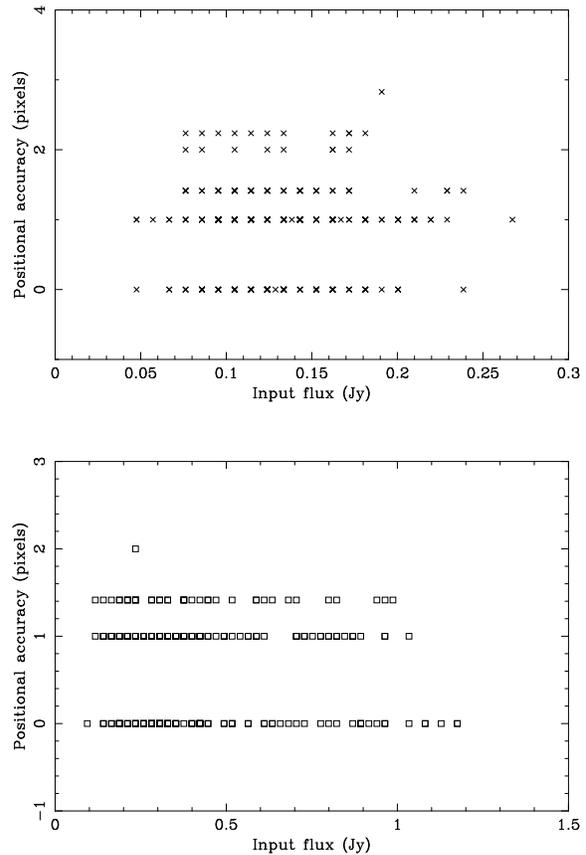

\begin{center}
\includegraphics[width=150pt,angle=270]{figure11a.ps}\vspace{20pt}
\includegraphics[width=150pt,angle=270]{figure11b.ps}
\end{center}
\caption[Positional accuracy of source detection]{Positional accuracy in source detection.  The quantisation here reflects the pixelised nature of the maps, where 1 pixel = 3\arcsec.  Upper and lower plots are as in Fig. \ref{fig:inp_det}.}
\label{fig:posacc_inp}
\end{figure}

The positional accuracy of the detections can also be quantified.  Fig. \ref{fig:posacc_inp} shows the positional accuracy of each detection against the input flux required for first detection of a source. It shows broad distributions for both the Perseus and Galactic centre regions.  In these graphs the positional accuracy is the absolute distance between the input and output positions, in units of 3\arcsec pixels.  The positional accuracy of the detections is extremely reliable with respect to the size of the beam. 

\subsection{Comparison with unfiltered maps}
\begin{figure}
\begin{center}
\includegraphics[width=150pt,angle=270]{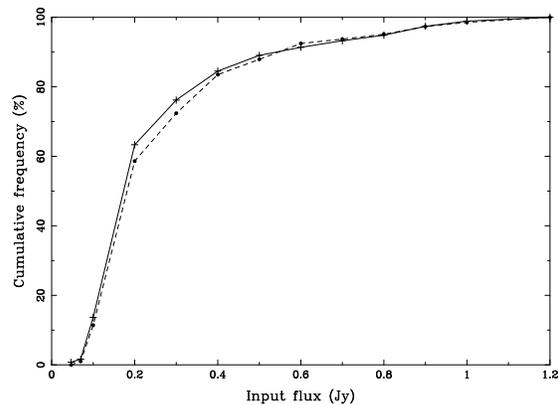}
\caption[Unfiltered simulation results I]{Unfiltered simulation results: cumulative counts.  The overall results for the filtered sources are shown by the solid line with crosses and the same results for unfiltered sources are shown by the dashed line with solid circles.}
\label{fig:nfsummary}
\end{center}
\end{figure}

\begin{figure}
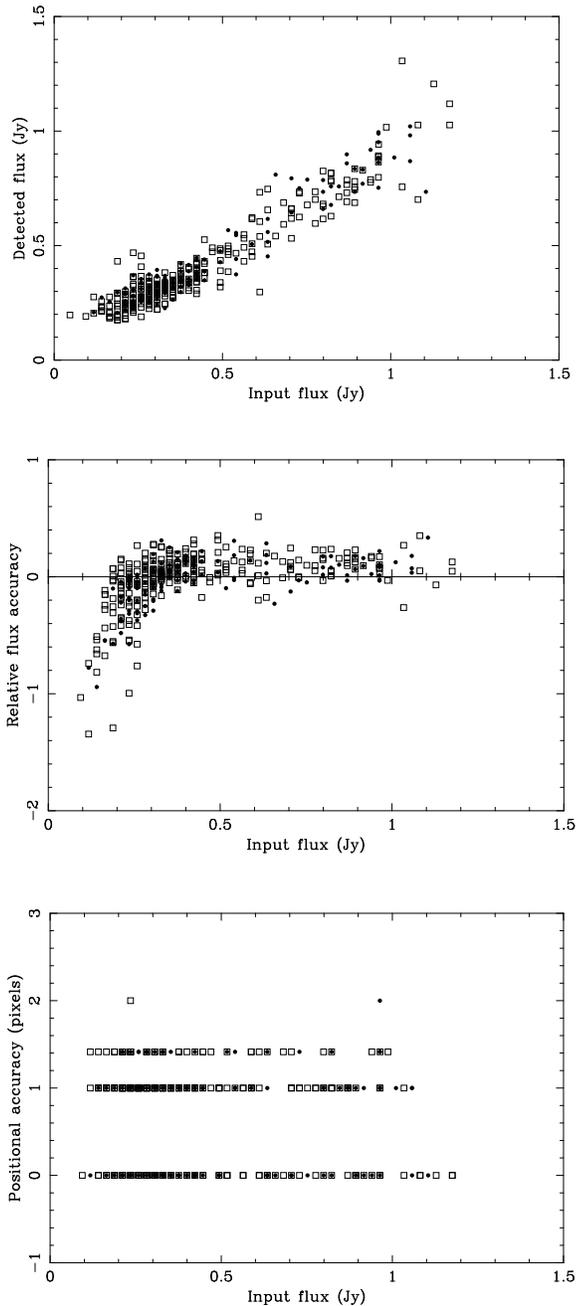

\begin{center}
\includegraphics[width=150pt,angle=270]{figure13a.ps}\vspace{20pt}
\includegraphics[width=150pt,angle=270]{figure13b.ps}\vspace{20pt}
\includegraphics[width=150pt,angle=270]{figure13c.ps}
\caption[Unfiltered simulation results for the Galactic centre]{Comparison of filtered and unfiltered results for the Galactic centre.  Filtered sources are shown as squares, as earlier.  Unfiltered sources are shown as solid circles.  Results for the Perseus region showed similar overlap but are not shown for brevity.}
\label{fig:nfsummarygc}
\end{center}
\end{figure}
To understand the impact of the filtering on the appearance of true point sources, the same simulations were run but without filtering the map at the recombination stage in the data reduction.  The same data analysis was completed, and Fig. \ref{fig:nfsummary} compares the cumulative counts for each version of the maps.  This figure demonstrates that, in the overall population, the sensitivity of the MHW technique is not affected by the filtering option.  Fig. \ref{fig:nfsummarygc} shows the detailed results for the Galactic centre, confirming the overlap in detection properties of filtered and unfiltered maps.  It is clear that the filtering procedure may create spurious extra sources but does not lose real ones.

\subsection[Loss of point sources]{Possible problems with MHW and SCUBA scan-maps: Loss of point sources}
\label{sec:losses}
During the analysis of the simulations, some strange behaviour in the detection history of individual sources became apparent:  a simulated source detectable at a lower input flux was sometimes not detected at the next higher input flux.  Also, real sources in the fields sometimes disappeared from the detection lists as the fluxes of the additional simulated sources increased.  Not every source nor every field was affected however.  It appears that two modes of `point source loss' were occurring.

\subsubsection{Mode A: Increasing $\chi^2$ as flux increases}

In this mode, point sources were not detected by the MHW routines because their $\chi^2$ values were too high.  The $\chi^2$ parameter (section \ref{sec:detpars}) is a measure of how similar the shape of a point source is to a Gaussian.  The limiting value of 4 used here was determined from previous simulations of \emph{Planck} detections.  Whilst the \emph{Planck} simulations were run for the same wavelength as these SCUBA observations, the simulated Galactic emission was of a general cirrus component on larger spatial scales, and so was not representative of the high-flux and variable structures in these maps.  The \emph{Planck} beam was also assumed to be perfectly Gaussian.  The beam map in Fig. \ref{fig:scan_calib_850} shows that the beam of the SCUBA scan-map mode is not a perfect Gaussian; in particular there are features related to the image reconstruction technique.  As the flux of a simulated source increases, imperfections in the beam become more important relative to the local background, increasing the $\chi^2$ value of the source and preventing its detection.  

To check this explanation, the simulations were repeated with two variations.  Firstly, a single point source was added to a map, rather than several at once, to check that the beam profiles of adjacent simulated sources were not interfering.  The detection history for a particular source was found to be identical whether it was added alone to a map or multiple sources were added at once, indicating that overlapping of sources was not an important factor in these simulations.  Secondly, the individual sources were replaced by perfect Gaussians of the same height and beamwidth.  In this case no loss of sources was observed at all.  These two results confirm that it is likely to be the imperfectly Gaussian beams of the individual scan-map point sources which causes their loss at higher fluxes.  

Having discovered this effect in the simulations, it now seems likely that this is the cause of the very weak detection of the very bright source gc6 in the initial maps (see section \ref{sec:detpars}).  This exceptionally bright source, which from the jiggle-map observations appears still to be a real point source, was only detected by the MHW routines with some relaxation of the detection parameters, to allow a higher $\chi^2$ value of 5.26.  Given its flux, its seems probable that its high $\chi^2$ value was due to the impact of the outer imperfections in the beam.  

In general this mode of point source loss affects only the brighter sources in SCUBA scan-maps.  A check by eye of all fields being surveyed identified any possible problem sources, and we are confident that no real bright point sources were lost in this manner, gc6 having been found.  The flux level at which sources are lost varied from position to position, though generally this loss mode only occurred at input fluxes $\ga 800$ mJy; it depended on the local features of the background around the source, since these determined when the imperfections of the beam become significant.

\subsubsection{Mode B: the contribution of point sources to $\sigma_{w_n}(R)$}

A second mode of point source loss was observed in the simulation results.  Unlike the previous mode, this could occur at both low and high flux levels, and affected the real point sources in the map, as well as the simulated sources.  

In this loss mode, sources fail the second detection requirement, that the signal-to-noise in wavelet space,  $D_w(R_{\rm{opt}})$, is $\geq 5$ which was also set from the \emph{Planck} simulations.  An implicit assumption within this formalism is that the point sources themselves are not affecting the value of $\sigma_{w_n}(R_{\rm{opt}})$, the noise in wavelet space.  $\sigma_{w_n}(R_{\rm{opt}})$ is calculated from the properties of the whole map, and ideally should be determined from a background without any point sources. In practice this is not plausible, as the Galactic centre and Perseus backgrounds are hard to model.  Instead, the point sources in a map \emph{can} contribute to the value of $\sigma_{w_n}(R)$ at high enough flux levels, and so affect their own chance of detection.  

Similar checks were carried out to test the behaviour of this mode of point source loss as for mode A.  The effect was reduced but did not completely disappear with perfectly Gaussian point sources.  Also the number of point sources added to a map did make a difference to the results.  This mode of point source loss is related therefore to the overall properties of the image map used, as well as to the properties of the sources.  In smaller maps the presence of bright point sources is more likely to make an impact on the value of $\sigma_{w_n}(R_{\rm{opt}})$, possibly to the extent that fainter point sources will not be detectable.  For regions where the background cannot be simulated, two ways to avoid this effect would be to either use larger maps, or to alter the detection requirements to lower $D_w(R_{\rm{opt}})$ values for maps which are either crowded or contain bright sources, in order to be sure of detecting the fainter sources.  The latter of course will increase the number of spurious detections as well though.  This mode of point source loss will only have affected the survey results for rare fields in which there were very bright point sources.  Since each field was checked by eye, we are again confident that no real sources were lost in this way.

\subsection{Summary of simulation results}

In general the simulations suggest that the survey carried out with the MHW algorithms has produced reliable results, and the small average errors in flux and position estimation are easily characterisable.  It appears that the filtering option used in the construction of the final images did not affect the detection of real point sources as a population.  Two mechanisms which may lead to non-detections of real point sources have been identified, but it seems unlikely that either strongly affected the result of this survey.

\section{Final results}

The final results of this paper are constraints put on the counts of SCUBA galaxies at bright fluxes.  This is somewhat simplified (!) by the fact that no extragalactic candidates were detected.  However the results of the simulations indicate the sensitivity of the source detection method used, and so were used to set appropriate upper limits on the possible population.

Poisson confidence limits for small numbers were published by \citet{gehrels}.  The upper limit corresponding to a 1-$\sigma$ Gaussian limit is 1.841.\footnote{The equivalent limits for 2- and 3-$\sigma$ are 2.303 and 2.996.  For a zero count there is no lower limit.}  For flux values in the range suggested by Fig. \ref{fig:area_sens}, the upper limit was calculated as ${1.841}/{A_{total}}$.  This resulted in the final error bars listed in Table \ref{counts_table}.  Below flux limits of approximately 200 mJy the Galactic centre survey is relatively insensitive, whereas the Perseus survey is complete above about 300 mJy.  Hence the 1-$\sigma$ upper limit on the 850-$\mu\rm{m}$ count is 53 per square degrees at 100 mJy and 2.9 per square degree at 50 mJy.

\begin{table*}
\caption[Final limits]{Final results.  Columns 2, 3 and 4 show the area over which the survey was sensitive to the flux limit in, respectively, the Perseus region, the Galactic centre and the two regions combined.  Column 5 gives the 1 $\sigma$ upper limit as calculated as ${1.841}/{\rm{A_{total}}}$.}
\label{counts_table}
\begin{center}
\begin{tabular}{ccccc}\hline
Flux limit (Jy) & $\rm{A_{pers}}$ (sq. deg.) & $\rm{A_{gc}}$ (sq. deg.) & $\rm{A_{total}}$ (sq. deg.) & $\sigma_{\rm{A}}$ (per sq. deg.)\\\hline
0.05 & 0.028 & 0.007 & 0.035 & 53 \\
0.07 & 0.070 & 0.007 & 0.077 & 24 \\
0.1 & 0.616 & 0.018 & 0.634 & 2.9 \\
0.15 & 1.930 & 0.107 & 2.037 & 0.9 \\
0.2 & 2.657 & 0.280 & 2.937 & 0.6 \\
0.3 & 2.797 & 0.736 & 3.533 & 0.52 \\
0.4 & 2.797 & 1.123 & 3.920 & 0.47 \\
0.5 & 2.797 & 1.333 & 4.130 & 0.45 \\
0.6 & 2.797 & 1.440 & 4.237 & 0.43 \\
0.7 & 2.797 & 1.528 & 4.325 & 0.43 \\
0.8 & 2.797 & 1.602 & 4.399 & 0.42 \\
0.9 & 2.797 & 1.727 & 4.524 & 0.41 \\
1.0 & 2.797 & 1.790 & 4.587 & 0.40 \\
1.2 & 2.797 & 1.841 & 4.638 & 0.40 \\\hline
\end{tabular}
\end{center}
\end{table*}

\begin{figure}
\begin{center}
\includegraphics[width=150pt,angle=270]{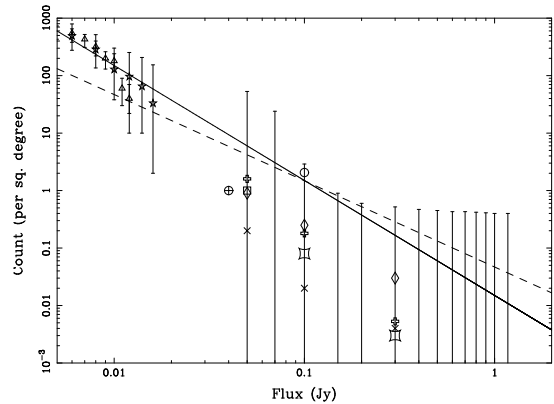}
\end{center}
\caption[Euclidean and modelling predictions for counts]{Euclidean and modelling predictions for counts.  The dashed line shows the highest possible Euclidean slope allowed by these data.  The solid line represents the slope of the count determined at fainter fluxes, as shown in Fig. 1. The large symbols at higher fluxes show predictions from six galaxy evolution models predicting bright counts at 850 $\mu \rm{m}$: \citealt{f10} (diamonds) ; \citealt{g9} (circle);  \citealt{g8} (crossed circle); \citealt{d6} (curved squares); \citealt{r7} (diagonal crosses); \citealt{p7} (square); \citealt{bsik} (open crosses).  The smaller symbols with error bars to the left show the results from the two brightest blank-field SCUBA surveys to date (triangles: Scott et al.\ 2002; stars: Borys et al.\ 2003)}
\label{fig:counts2}
\end{figure}

In Fig. \ref{fig:counts2}, the upper limits in table \ref{counts_table} are plotted, with two lines.  The dashed line is a Euclidean slope ($\propto S^{-3/2}$), with the highest intercept consistent with \emph{all} the upper limit error bars: at $S = 130\, \rm{mJy}$, the count of galaxies is 1 per square degree.  The strongest constraints on this normalisation are the error bars at 150 -- 200 mJy, where the combined bright survey becomes sensitive over more than 50\, per cent of the available area (see Table \ref{counts_table}).  The solid line represents the power law fitted in Fig. \ref{fig:blankcounts} to the fainter SCUBA galaxy counts.  Also plotted are the results of the two brightest blank-field SCUBA surveys (\citealt{s17,borys_super}, see table 1 for details).

The count of galaxies at fluxes lower than 10 mJy is clearly super-Euclidean.  However both the 8 mJy survey and HDF super-map results appear to show a slight turnover at their brightest fluxes, $\sim$ 10 -- 15 mJy.  Whilst these results are of course subject to larger uncertainties than at lower fluxes, the counts at these bright fluxes are more consistent with the Euclidean slope fitted to the upper limits derived in this paper.  This supports the idea that the steep turnover seen in the $\sim$ 15 mJy counts is real and continues to brighter fluxes.

This result also may be consistent with a suggestion by \citet{l6} that a \emph{minority} of the blank-field SCUBA sources, even at fluxes as low as 2 mJy, might be local, very cold, dark clouds at distances of around 100 pc.  Such sources would be expected to show a Euclidean trend at fluxes brighter than about 10 mJy, although recent confirmation of cosmological redshifts for the majority of radio-selected SCUBA-detected objects \citep{scott's_redshifts,scott2} confirm that any such Galactic population is not numerically significant.  The Euclidean slope plotted here corresponds to about 20 per cent of the count found by \citet{s17} at 8mJy, and nearly 90 per cent at 12 mJy.

Also plotted on Fig. \ref{fig:counts2} are the predictions for the bright 850-$\mu \rm{m}$ count from various galaxy evolution models which give bright count predictions.  These are all consistent with the limits found from the Galactic centre and Perseus surveys.

\section{Conclusions}

In this paper an application of the Mexican Hat Wavelet to detect SCUBA galaxies was presented.  Bright candidates were detected in scan-maps of the Galactic centre and the Perseus star-forming region.  Follow-up continuum and spectral line observations suggested that none of the 21 initial candidates were extragalactic; furthermore, only three appeared to be genuine point sources.  Re-reducing the scan-map data without filtering its Fourier spectrum removed the spurious point source candidates.

Simulations of the MHW on the Galactic centre and Perseus fields demonstrated that the technique would have detected point sources flux over the flux range 50 -- 1000 mJy.  The flux and positional accuracies of recovered sources were shown to be high.  A problem involving the loss of point sources was revealed, but overall the MHW technique has proved robust, particularly in the presence of very complicated backgrounds.  The two key advantages of the MHW techniques are i) the quantitative evaluation of both the shape and flux of the source, and ii) that no assumptions are required about the characteristics of the noise in an image, which is often an implicit feature of other routines.

For maps made with SCUBA, the MHW technique is particularly appropriate when the beam is complicated, since the central Gaussian feature to which the MHW routines are sensitive will still remain.  The MHW is also likely to be the most robust technique in maps with high or varying backgrounds.  This work has demonstrated that point sources can be reliably located and measured in SCUBA scan-maps when a suitable detection algorithm is used, and that the scan-mapping mode should now be considered a possible tool for extragalactic submillimetre surveys.  

Using the results of simulations, upper limits have been established for the count of SCUBA galaxies at fluxes $>$ 50 mJy.  The Euclidean slope then consistent with these upper limits approaches the previous brightest counts at fluxes $\sim$ 15 mJy, suggesting that the turnover seen in the brightest blank-field surveys is real and continues to very bright fluxes.

\section*{Acknowledgements}
We thank Kirsten Kraiberg Knudsen for useful comments and discussions and Tim Jenness for comments and advice regarding \textsc{Surf}.  We also thank an anonymous referee.  AWB acknowledges support from NSF grant AST-0205937, the Research Corporation and the Alfred P. Sloan foundation.  The JCMT is operated by the Joint Astronomy Centre in Hilo, Hawaii
   on behalf of the Particle Physics and Astronomy
   Research Council in the United Kingdom, the National Research Council
   of Canada and The Netherlands Organization for Scientific Research. The
   authors acknowledge the data analysis software provided by the Starlink
   project which is run by the CCLRC on behalf of PPARC.

\end{document}